\documentclass[submission,copyright,creativecommons]{eptcs}
\providecommand{\event}{ACL2 2020} 
\usepackage{breakurl}             
\usepackage{underscore}           
\usepackage{comment}
\usepackage{fancyvrb}

\usepackage{float}
\floatstyle{boxed}
\restylefloat{figure}

\title{Computing and Proving Well-founded Orderings through Finite Abstractions}
\author{Rob Sumners
\institute{Centaur Technology}
\email{rsumners@centtech.com}
}
\def\titlerunning{Computing and Proving Well-founded Orderings through Finite Abstractions}
\def\authorrunning{Rob Sumners}
\begin{document}
\maketitle

\begin{abstract}
  A common technique for checking properties of complex state machines is to build a finite
  abstraction then check the property on the abstract system --- where a passing check on the
  abstract system is only transferred to the original system if the abstraction is proven to be
  representative. This approach does require the derivation or definition of the finite
  abstraction, but can avoid the need for complex invariant definition. For our work in checking
  progress of memory transactions in microprocessors, we need to prove that transactions in complex
  state machines always make progress to completion. As a part of this effort, we developed a
  process for computing a finite abstract graph of the target state machine along with annotations
  on whether certain measures decrease or not on arcs in the abstract graph. We then iteratively
  divide the abstract graph by splitting into strongly connected components and then building a
  measure for every node in the abstract graph which is ensured to be reducing on every transition
  of the original system guaranteeing progress. For finite state target systems (e.g. hardware
  designs), we present approaches for extracting the abstract graph efficiently using incremental
  SAT through GL and then the application of our process to check for progress. We present an
  implementation of the Bakery algorithm as an example application.
\end{abstract}

\section{Introduction} \label{sec:intro}

In order to admit a recursive function to ACL2, the user must prove that the function terminates by
showing that a function of the inputs exists which returns an ordinal (recognized by \texttt{o-p})
and which strictly decreases (by \texttt{o<}) on every recursive call of the function. The
epsilon-$0$ ordinals recognized by \texttt{o-p} and ordered by \texttt{o<} are axiomatized in this
way to be well-founded in ACL2. Our goal in this work is to present a new way to prove that certain
relations are well-founded. Referring to Figure~\ref{fig:wellfound} and given a relation \texttt{(r
  x y)}, we produce a measure function \texttt{(m x)} and proofs of the properties
\texttt{m-is-an-ordinal} and {\tt m-is-o<-when-r}. This entails that the given relation \texttt{r}
is well-founded.

\begin{figure}[!htbp]
\scriptsize
\begin{verbatim}
(encapsulate (((r * *) => *) ((m *) => *))
     ...
  (defthm m-is-an-ordinal (o-p (m x)))
  (defthm m-is-o<-when-r  (implies (r x y) (o< (m y) (m x)))))
  
(defchoose choose-r (y) (x) (r x y))

(defun seq-r (x)  ;; any sequence of objects related by R must terminate
  (declare (xargs :measure (m x)))
  (if (r x (choose-r x)) (seq-r (choose-r x)) x))
\end{verbatim}
\normalsize
\caption{Proving a relation is well-founded}
\label{fig:wellfound}
\end{figure}

In order to admit recursive functions, ACL2 has built-in heuristics for guessing appropriate
measures and attempts to prove them. These heuristics often work for functions with common recursive
patterns with user specification of measures covering the remaining cases. The theorem prover
ACL2s~\cite{acl2s} (the sedan) has a built-in procedure which builds so-called Calling Context Graphs
(or CCGs)~\cite{ccg} and checks that there are no infinite paths through the CCG such that some
measure doesn't decrease infinitely often while never increasing. The CCG checker in ACL2s
significantly increases the number of functions which can be admitted without user specification of
measures. Our work shares some similarities at a high level to the work on CCGs in ACL2s but the
approach and target applications are quite different --- we will cover these differences in greater
detail in Section~\ref{sec:concl}.

Our primary focus is proving well-founded relations \texttt{(r x y)} derived from software and
hardware systems comprised of interacting state machines. In particular, for the work presented in
this paper, we focus on systems defined as the composition of finite state machines. We present a
procedure in this paper which takes the definition of \texttt{(r x y)}, a finite domain
specification for \texttt{x} and \texttt{y} and a mapping from the concrete finite domain for
\texttt{x} and \texttt{y} to an abstract domain. The procedure leverages existing bit-blasting tools
in ACL2 to construct the abstraction of \texttt{r}, builds an abstract measure descriptor, and then
translates this back to a proven measure on the concrete domain. In Section~\ref{sec:bakery}, we
cover a version of the Bakery algorithm which will be the primary example for this paper. In
Section~\ref{sec:overview}, we present an overview of the procedure for generating and proving the
needed measures and in Sections~\ref{sec:gl},~\ref{sec:scc},~\ref{sec:proof}, we go into the details
of the steps of the procedure along with demonstration via application to the example. We conclude
the paper in Section~\ref{sec:concl} with a discussion of related work and future considerations.


\section{Example: Bakery Algorithm} \label{sec:bakery}

We will use a finite version of the Bakery algorithm as an example application throughout this
paper.  The Bakery algorithm was developed by Lamport~\cite{Lamport} as a solution to mutual
exclusion with the additional assurance that every task would eventually gain access to its
exclusive section. The Bakery algorithm has also been a focus of previous ACL2 proof
efforts~\cite{RaySumners, somenerd}.

The Bakery algorithm operates by allowing each process that wants exclusive access to first choose a
number to get its position in line and then later compares the number against the numbers chosen by
the other processes to determine who should have access to the exclusive section. The Bakery
algorithm definition we will use for presenting this work is defined in Figure~\ref{fig:bakeimpl}
where the macros \texttt{(tr+ ..)} and \texttt{(sh+ ..)} are shorthand for functions which update the
specified fields of the \texttt{bake-tr-p} and \texttt{bake-sh-p} data structures.

The function \texttt{(bake-tr-next a sh)} takes a local bakery transaction state ``\texttt{a}'' and
a shared state ``\texttt{sh}'' and updates the local bakery state. The function
\texttt{(bake-sh-next sh a)} takes the same ``\texttt{sh}'' and ``\texttt{a}'' and produces the
updated shared state (a single variable \texttt{sh.max} storing the next position in line). The
function \texttt{(bake-tr-blok a b)} defines a blocking relation which denotes when one bakery
process ``\texttt{a}'' is blocked by another bakery process ``\texttt{b}''.

Each task will start in program location $0$ in which it starts its \texttt{a.choosing}
phase. During the \texttt{a.choosing} phase, the task will grab the current shared max variable
\texttt{sh.max} and then set its own position \texttt{a.pos} to be $1$ more than \texttt{sh.max} ---
possibly wrapping around to a position of $0$. If we do wrap the position around to $0$, then the
local Bakery process will cycle through the other Bakery processes that are completing to allow them
to flush out before proceeding. After this check, the process will perform an atomic
compare-and-update at program location (or {\tt a.loc}) $6$ to the shared \texttt{sh.max} variable
and ends its \texttt{a.choosing} phase.

After the \texttt{a.choosing} phase, the process at \texttt{a.loc} $7$ will enter another loop to
check if it can proceed to the critical section. In locations $8$, $9$, and $10$, the process checks
if the current process we are checking (at index \texttt{a.loop}) is either still choosing a
position in line or is ahead of us in line (with ties broken by checks on the order of process indexes
at location $10$). Finally, after the process enters and exits the critical section at location
$13$, the process will decrement the \texttt{a.runs} outer loop count and either branch back to
location $0$ or complete and set \texttt{a.done} at location $17$.

\begin{figure}[!htbp]
\scriptsize
\begin{verbatim}
(define bake-tr-next ((a bake-tr-p) (sh bake-sh-p))
  (b* (((bake-tr a) a) ((bake-sh sh) sh))
    (case a.loc
      (0  (tr+ a :loc 1
                 :choosing t))
      (1  (tr+ a :loc 2 
                 :temp sh.max))
      (2  (tr+ a :loc 3 
                 :pos (ctr-1+ a.temp) ;; can wrap to 0
                 :loop (loop-start)))
      (3  (tr+ a :loc 4))  ;; possibly blocked here
      (4  (tr+ a :loc 5 
                 :loop (1- a.loop)))
      (5  (tr+ a :loc (if (= a.loop 0) 6 3)
                 :pos-valid (= a.loop 0)))
      (6  (tr+ a :loc 7))  ;; update shared variable
      (7  (tr+ a :loc 8 
                 :choosing nil
                 :loop (loop-start)))
      (8  (tr+ a :loc 9))  ;; possibly blocked here
      (9  (tr+ a :loc 10)) ;; possibly blocked here
      (10 (tr+ a :loc 11)) ;; possibly blocked here
      (11 (tr+ a :loc 12 
                 :loop (1- a.loop)))
      (12 (tr+ a :loc (if (= a.loop 0) 13 8)))
      (13 (tr+ a :loc 14   ;; critical section
                 :pos-valid nil))
      (14 (tr+ a :loc 15 
                 :runs (1- a.runs)))
      (15 (tr+ a :loc (if (= a.runs 0) 16 0)))
      (t  (tr+ a :loc 17   ;; process is done
                 :done t)))))

(define bake-sh-next ((sh bake-sh-p) (a bake-tr-p))
  (b* (((bake-tr a) a) ((bake-sh sh) sh))
    (case a.loc
      (6 (if (not (ctr-> sh.max a.temp)) ;; careful on wrap to 0
             (sh+ sh :max a.pos)
           sh))
      (t sh))))

(define bake-tr-blok ((a bake-tr-p) (b bake-tr-p))
  (b* (((bake-tr a) a) ((bake-tr b) b))
    (and (= a.loop b.ndx)
         (case a.loc
           (3  (and (= a.pos 0) b.pos-valid))
           (8  (and (not (= b.pos 0)) b.choosing))
           (9  (and b.pos-valid (< b.pos a.pos)))
           (10 (and b.pos-valid (= b.pos a.pos)
                    (< b.ndx a.ndx)))))))
\end{verbatim}
\normalsize
\caption{Bakery Process Definitions}
\label{fig:bakeimpl}
\end{figure}

Our goal is to prove that a system comprised of some number of bakery processes updating
asynchronously will eventually reach a state where all bakery processes are done. This is codified
by admitting the function \texttt{(bake-run st)} defined in Figure~\ref{fig:bakerun}. The {\tt
  bake-run} function takes a state \texttt{st} consisting of a list \texttt{st.trs} of transaction states
(one for each bakery process) and a shared variable state \texttt{st.sh}. The function checks if all
bakery processes are done (i.e.\ \texttt{(bake-all-done st.trs)}) and simply returns if so. Otherwise,
the function chooses a process which is ready and updates the state for that process along with the shared
state and recurs. The function \texttt{(choose-ready st.trs st.sh oracle)} is constrained (via {\tt
  encapsulate}) to return an index for a bakery process state which is not done and is not blocked
by any other process state (via \texttt{bake-tr-blok}). Given the function \texttt{choose-ready} is
constrained to represent any legal input selection, then \texttt{bake-run} represents all legal bakery
runs and its termination ensures that all runs end with all bakery processes done. We note that this
does not prove the Bakery algorithm ensures mutual exclusion and it does not prove that the Bakery
algorithm avoids livelock or starvation --- these issues were covered in ~\cite{somenerd} but require
more complex specifications involving infinite runs which are not closed-form in ACL2. The work
presented in this paper to generate proven measures for well-founded relations has been applied to
the more complete proof framework presented in ~\cite{somenerd}.

\begin{figure}[!htbp]
\scriptsize
\begin{verbatim}
(define bake-run (st orcl)
  (b* (((bake-st st) st))
    (if (bake-all-done st.trs) 
        st
      (b* ((n   (choose-ready st.trs st.sh orcl))
           (a   (nth n st.trs))
           (trs (update-nth n (bake-tr-next a st.sh) st.trs))
           (sh  (bake-sh-next st.sh a)))
        (bake-run (make-bake-st :trs trs :sh sh) (next-oracle orcl))))))
\end{verbatim}
\normalsize
\caption{Bakery System Run Function}
\label{fig:bakerun}
\end{figure}

\section{Overview} \label{sec:overview}

Our goal is to define \texttt{bake-run} and admit it by proving its termination. In support of this
goal, we need to prove that two relations are well-founded orderings. First, we need to build a
measure showing that on updates with \texttt{bake-tr-next}, each bakery process makes progress to a
done state. The other measure we need to define and prove is a little more subtle. The function
\texttt{choose-ready} must return a process index with a state which is not done and is not
blocked. In the case of a deadlock between some number of process states (a cycle of the
\texttt{bake-tr-blok} relation), choosing an unblocked process may not be possible. We need to
build a measure showing that no blocking cycles exist between states; this measure will allow us
to define a function which always finds an unblocked process state. In each of these cases, we begin
with a relation \texttt{(r x y)} that we want to show is well-founded requiring the definition of a
measure \texttt{(m x)} which preserves the properties \texttt{m-is-an-ordinal} and
\texttt{m-is-o<-when-r} from Figure~\ref{fig:wellfound}.

\begin{figure}[!htbp]
\begin{enumerate}
\item {\bf Bakery algorithm definition} :
  \begin{itemize}
  \item file: \texttt{bakery.lisp}
    \begin{itemize}
    \item Defines the types and functions for the Bakery algorithm process states.
    \item Defines \texttt{GL} shape specifiers and theorems to connect \texttt{GL} shapes to the types.
    \end{itemize}
  \end{itemize}
\item {\bf Building abstract models with \texttt{GL} and Incremental SAT} :
  \begin{itemize}
  \item file: \texttt{gl-fin-set.lisp}
    \begin{itemize}
    \item Defines function for computing possible values for a term with finite variable bindings.
    \item Uses \texttt{GL} functions to translate the term to CNF formula.
    \item Uses \texttt{IPASIR} (Incremental SAT) to efficiently find a number of values resulting from the given term.
    \end{itemize}
  \item file: \texttt{gen-models.lisp}
    \begin{itemize}
    \item Defines reachable abstract graph builders using the functions from \texttt{gl-fin-set.lisp}.
    \item Defines functions to tag these graphs with which orderings decrease along the arcs.
    \end{itemize}
  \end{itemize}
\item {\bf Building measure descriptors from abstract models} :
  \begin{itemize}
  \item file: \texttt{cycle-check.lisp}
    \begin{itemize}
    \item Defines graph algorithms for processing an abstract graph with ordering tags
      which produces either a cycle in the graph for a failing case or a measure descriptor in
      a passing case.
    \end{itemize}
  \item file: \texttt{bake-models.lisp}
    \begin{itemize}
    \item Calls abstract model generation from \texttt{gen-models.lisp} to build abstract models for the Bakery example.
    \item Calls graph algorithms from \texttt{cycle-check.lisp} to build measure descriptors for the Bakery example.
    \end{itemize}
  \end{itemize}
\item {\bf Building proven measures from measure descriptors} :
  \begin{itemize}
  \item file: \texttt{wfo-thry.lisp}
    \begin{itemize}
    \item Builds a theory relating properties of the abstract model and generated measure descriptor to proving a measure.
    \item Uses theory of natural number lists and lists of natural lists defined in \texttt{bounded-ords.lisp}.
    \end{itemize}
  \item file: \texttt{bake-proofs.lisp}
    \begin{itemize}
    \item Instantiates theory from \texttt{wfo-thry.lisp} with abstract models and measure descriptors from \texttt{bake-models} to build proofs of the needed measures.
    \item Also proves some auxiliary properties needed for Bakery proof.
    \end{itemize}
  \end{itemize}
\item {\bf Admitting \texttt{bake-run} using proven measures} :
  \begin{itemize}
  \item file: \texttt{top.lisp}
    \begin{itemize}
    \item Uses proven measures from \texttt{bake-proofs.lisp} to define \texttt{choose-ready} and build a measure to admit \texttt{bake-run}
    \end{itemize}
  \end{itemize}
\end{enumerate}
\caption{Overview of the Procedure}
\label{fig:overview}
\end{figure}

A standard ACL2 proof that these relations are well-founded would not only require defining a
decreasing measure \texttt{(m x)}, but in addition and invariably, invariant properties of the
reachable process states. In order to prove these invariants, the user will need to strengthen the
invariant definitions to be inductive. For complex systems, the definition of inductive invariants
can be prohibitively expensive. In addition, inductive invariant definitions are fragile and require
maintenance when system definition changes. The same is also true of defining measures for proving
well-founded relations in complex systems --- these measure definitions can also be extensive and
brittle to changes in system definition. Our goal is to build a procedure which allows the
generation of inductive invariants as well as decreasing measures which prove our target relations
to be well-founded. Computing invariant and measure definitions not only has the benefits of
requiring less human definition and tracking design change, but can also provide more direct debug
output from failed attempts. We provide an overview of the steps in our procedure (as well as
pointers to where these steps are defined in the supporting materials) in
Figure~\ref{fig:overview}. We cover each of these steps in greater detail in the remaining sections
of the paper as well as covering their application to our Bakery algorithm example.

\section{Building Models with GL and Incremental SAT} \label{sec:gl}

The tool \texttt{GL}~\cite{gl} is an extension to ACL2 (primarily an untrusted clause processor)
which targets proving theorems on finite domains by translating the theorems to boolean formulas
using symbolic simulation and then checking the boolean formulas through BDDs or SAT with boolean AIG
transformations and simplifications. There are different ways to direct \texttt{GL} to translate a
term to a boolean formula, but the most basic form is to take a \texttt{hyp} term and \texttt{concl}
term along with a shape specification \texttt{g-bindings} for the free variables in the \texttt{hyp}
and \texttt{concl} terms. \texttt{GL} uses the shape specification to provide a symbolic value for
the free variables and then symbolically evaluates the \texttt{concl} and checks if the resulting
boolean formula is valid. If the boolean check passes, then \texttt{GL} checks that the \texttt{hyp}
implies the constraints specified by the shape specification. The first step in our procedure uses
the setup for \texttt{GL} but instead of proving that a term is always true, we will instead compute the
set of values that a term can return under evaluation with variable bindings consistent with
\texttt{g-bindings}.

We define the function \texttt{(compute-finite-values trm hyp g-b num)} which takes terms
\texttt{trm} and \texttt{hyp}, shape specification \texttt{g-b}, and natural number \texttt{num} and
attempts to return (up-to \texttt{num}) values from the set of possible values for \texttt{trm} under
the assumption of \texttt{hyp} with free variables consistent with \texttt{g-b}. The function
\texttt{compute-finite-values} will also return a boolean \texttt{is-total} which is true if the
list of values returned is the entire set of possible values. We use the \texttt{GL} symbolic
evaluation functions to provide a translation from terms (with shape-specifications) to boolean CNF
formula and then iterate through the set of possible boolean values discovered using incremental SAT
via the \texttt{IPASIR} library~\cite{ipasir}. The resulting boolean valuations from repeated IPASIR
tests are then translated back to ACL2 objects and returned.

This inner IPASIR Incremental SAT loop begins with installation of the CNF formula (from the GL
translation) into the IPASIR clause database. The literals in the CNF formula corresponding to the
output of \texttt{trm} are also recorded. Then, within each iteration of the loop, we first call
IPASIR to find a satisfying assignment. If it is unsatisfiable then there are no more values and we
return. Otherwise, we retrieve the boolean values for the \texttt{trm} literals from IPASIR and add
this boolean valuation to our accumulated return set. We then add the negation of the equality of
the \texttt{trm} literals to the retrieved boolean values as a new clause in the IPASIR database and
iterate through the loop. We terminate the loop by either reaching an unsatisfiable IPASIR instance
or exceeding the user-specified maximum number \texttt{num} of values. The chief benefit of using
incremental SAT is the amortization of the translation and installation of the CNF formula along
with (and more importantly) the incremental benefits of any learned clauses that the SAT solver
determines through each iteration of this loop. \begin{footnote}{We note that independent of the
    work presented in this paper, a new revision of \texttt{GL}, named \texttt{FGL}, was added to
    ACL2 and integrates incremental SAT in addition to other features. In particular,
    \texttt{FGL} makes it much easier and direct to define exploration functions like
    \texttt{compute-finite-values} using rewrite rules, but we decided to present the approach in
    \texttt{GL} given that \texttt{GL} is more familiar to the ACL2 community at the time of the
    writing of this paper.}\end{footnote}

\begin{figure}[!h]
\scriptsize
\begin{verbatim}
(define bake-rank-map ((a bake-tr-p))
  (b* (((bake-tr a) a))
    `((:loc    ,a.loc)
      (:done   ,a.done)
      (:loop=0 ,(equal a.loop 0))
      (:runs=0 ,(equal a.runs 0))
      (:inv    ,(and (>= a.loop 0)
                     (>= a.runs 0))))))

(defconsts (*bake-rank-reach* state)
  (comp-map-reach :init-hyp `t
                  :init-trm `(bake-rank-map (bake-tr-init n r))
                  :step-hyp `(and (equal (bake-rank-map a) ,*src-var*)
                                  (not (bake-tr-done a)))
                  :step-trm `(bake-rank-map (bake-tr-next a sh))))
\end{verbatim}
\normalsize
\caption{Bakery state mapping and reachable graph construction}
\label{fig:rankmap}
\end{figure}

Using the \texttt{compute-finite-values} function, we construct an abstract graph from the concrete
system definition. The function \texttt{(comp-map-reach init-hyp init-trm step-hyp step-trm)} in
\texttt{gen-models.lisp} builds the reachable graph beginning with the set of values from
\texttt{init-trm} and iteratively reached by steps in \texttt{step-trm}. Returning to the Bakery
example, Figure~\ref{fig:rankmap} defines the mapping \texttt{(bake-rank-map a)} taking a bakery
process state and returning the state information needed to build a measure of progress to a done
state --- or, intuitively, the mapping \texttt{bake-rank-map} includes enough state information to
ensure a bakery process makes progress to a done state. This includes the current location
\texttt{a.loc}, whether or not we are done \texttt{a.done}, whether or not the inner loop or outer
loop variables have counted down to $0$, and then a predicate ensuring that the \texttt{a.loop} and
\texttt{a.runs} counters are natural. This last \texttt{:inv} predicate field is actually an
inductive invariant attached to the abstract state and we include it in the abstract state to
effectively prove and use this inductive invariant during the building of the abstract graph and
later, when we add ordering information to the graph.

The \texttt{comp-map-reach} function in Figure~\ref{fig:rankmap} builds the abstract reachable graph
by setting up calls to \texttt{compute-finite-values}. It first calls \texttt{compute-finite-values}
to return the set of values for \texttt{(bake-rank-map (bake-tr-init n r))} where \texttt{n} is
defined to be an index for the process state and \texttt{r} defines the number of runs or number of
iterations of the outer \texttt{bake-tr-next} loop. The function \texttt{comp-map-reach} builds a shape
specification based on the variables in the terms and then computes the initial states. The function
\texttt{comp-map-reach} will then iterate by computing the values for the step term
\texttt{(bake-rank-map (bake-tr-next a sh))} at each node with the hypothesis \texttt{(equal
  (bake-rank-map a) ,*src-var*)} --- during each step, the special variable \texttt{*src-var*} is
bound to the value of the current node in the reachable graph exploration (i.e.\ a reached result of
\texttt{bake-rank-map}).

\begin{figure}[!htbp]
\scriptsize
\begin{verbatim}
ACL2 !>(strip-cars *bake-rank-reach*)
(((:LOC 0)  (:DONE NIL) (:LOOP=0 T)   (:RUNS=0 NIL) (:INV T))
 ((:LOC 1)  (:DONE NIL) (:LOOP=0 T)   (:RUNS=0 NIL) (:INV T))
 ((:LOC 2)  (:DONE NIL) (:LOOP=0 T)   (:RUNS=0 NIL) (:INV T))
 ((:LOC 3)  (:DONE NIL) (:LOOP=0 NIL) (:RUNS=0 NIL) (:INV T))
 ((:LOC 4)  (:DONE NIL) (:LOOP=0 NIL) (:RUNS=0 NIL) (:INV T))
 ((:LOC 5)  (:DONE NIL) (:LOOP=0 NIL) (:RUNS=0 NIL) (:INV T))
 ((:LOC 5)  (:DONE NIL) (:LOOP=0 T)   (:RUNS=0 NIL) (:INV T))
 ((:LOC 6)  (:DONE NIL) (:LOOP=0 T)   (:RUNS=0 NIL) (:INV T))
 ((:LOC 7)  (:DONE NIL) (:LOOP=0 T)   (:RUNS=0 NIL) (:INV T))
 ((:LOC 8)  (:DONE NIL) (:LOOP=0 NIL) (:RUNS=0 NIL) (:INV T))
 ((:LOC 9)  (:DONE NIL) (:LOOP=0 NIL) (:RUNS=0 NIL) (:INV T))
 ((:LOC 10) (:DONE NIL) (:LOOP=0 NIL) (:RUNS=0 NIL) (:INV T))
 ((:LOC 11) (:DONE NIL) (:LOOP=0 NIL) (:RUNS=0 NIL) (:INV T))
 ((:LOC 12) (:DONE NIL) (:LOOP=0 NIL) (:RUNS=0 NIL) (:INV T))
 ((:LOC 12) (:DONE NIL) (:LOOP=0 T)   (:RUNS=0 NIL) (:INV T))
 ((:LOC 13) (:DONE NIL) (:LOOP=0 T)   (:RUNS=0 NIL) (:INV T))
 ((:LOC 14) (:DONE NIL) (:LOOP=0 T)   (:RUNS=0 NIL) (:INV T))
 ((:LOC 15) (:DONE NIL) (:LOOP=0 T)   (:RUNS=0 NIL) (:INV T))
 ((:LOC 15) (:DONE NIL) (:LOOP=0 T)   (:RUNS=0 T)   (:INV T))
 ((:LOC 16) (:DONE NIL) (:LOOP=0 T)   (:RUNS=0 T)   (:INV T))
 ((:LOC 17) (:DONE T)   (:LOOP=0 T)   (:RUNS=0 T)   (:INV T)))
\end{verbatim}
\normalsize
\caption{Nodes in reachable abstract graph from \texttt{comp-map-reach}}
\label{fig:reachnodes}
\end{figure}

The result of \texttt{comp-map-reach} is a graph defined as an alist where each pair in the alist
associates a node to a list of nodes which form the directed arcs --- the nodes are the results of
\texttt{bake-rank-map} for reachable bakery process states. The nodes in the
\texttt{*bake-rank-reach*} graph are included in Figure~\ref{fig:reachnodes}. There are $21$ nodes
consisting of $1$ node per location and $2$ nodes for locations $5$, $12$, and $15$. The extra nodes
for these locations is due to a split based on whether \texttt{a.loop} is equal to $0$ in locations $5$
and $12$ and whether \texttt{a.runs} is equal to $0$ in location $15$.

\begin{figure}[!htbp]
\scriptsize
\begin{verbatim}
(define bake-rank-ord ((a bake-tr-p) (o ord-p))
  (b* (((bake-tr a) a))
    (cond ((eq o 'runs) (nfix a.runs))
          ((eq o 'loop) (nfix a.loop))
          (t 0))))

(defconsts (*bake-rank-ord-graph* state)
  (comp-map-order :reach    *bake-rank-reach*
                  :ordr-hyp `(and (equal (bake-rank-map a) ,*src-var*)
                                  (equal (bake-rank-map b) ,*dst-var*)
                                  (equal (bake-tr-next a sh) b))
                  :ordr-trms (make-ordr-trms *bake-rank-ords* 
                                             'bake-rank-ord 'a 'b)))
\end{verbatim}
\normalsize
\caption{Bakery component measures and ordering tag construction}
\label{fig:ordmap}
\end{figure}

The reachable abstract graph for \texttt{bake-rank-map} is not sufficient to build a measure of
progress to a done state --- there are 3 backward arcs at locations $5$, $12$, and $15$. We could
solve this by adding the full values for \texttt{a.loop} and \texttt{a.runs} to
\texttt{bake-rank-map} but this would dramatically increase the number of nodes in the resulting
abstract graph and is clearly not viable in general. The better approach is to tag arcs in the
abstract graph with whether or not certain measures strictly decrease or possibly increase. This
ordering information is defined by the function \texttt{(bake-rank-ord a o)} and added via the
function \texttt{comp-map-order} in Figure~\ref{fig:ordmap}. The function \texttt{bake-rank-ord}
takes a bakery state and symbol identifying a component measure and returns the measure value
(natural values in this case but in general can be a list of natural numbers). The function
\texttt{comp-map-order} takes the reachable abstract graph and computes (using
\texttt{compute-finite-values}) tags for each arc in the reachable graph encoding whether the
specified component measure is either strictly-decreasing, not-increasing, or possibly-increasing
along that arc. The \texttt{runs} measure strictly decreases on the arcs from the node at location
$14$ to the nodes at location $15$ and is non-increasing on all arcs. The \texttt{loop} measure
strictly decreases on arcs $4$ to $5$ and from $11$ to $12$, increases on arcs $2$ to $3$ and $7$ to
$8$ and is non-increasing on all other arcs. This tagged reachable graph is used in the next section
to compute a measure descriptor covering the concrete relation used to build the graph --- in this
case, the next-state bakery function \texttt{bake-tr-next}.

\section{Building Measures with SCC decomposition} \label{sec:scc}

In the previous section, we used \texttt{GL} and \texttt{IPASIR} to construct abstract reachable
graphs with arcs tagged based on which component measures decreased or increased. The next step in
our procedure is to use an algorithm based on the decomposition of strongly connected components (SCCs) to
build an object describing how to build a full measure across the concrete relation represented by
the abstract graph. The algorithm consists of two alternating phases operating on subgraphs of the
original graph (starting with the original graph itself) and produces a mapping of the nodes in the
graph to a measure descriptor which is a list comprised of symbols (representing a component
measure) or natural numbers.

\begin{itemize}
\item if the current subgraph is an SCC:
  \begin{enumerate}
  \item search for a component measure which never increases and decreases at least once.
  \item if no such component measure is found, then find the minimal non-decreasing cycle and fail.
  \item otherwise, remove the component measure's decreasing arcs from the graph and recur.
  \item \texttt{cons} the component measure onto the measure descriptors from the recursive calls.
  \end{enumerate}
\item otherwise:
  \begin{enumerate}
  \item partition the graph into SCCs using a standard algorithm.
  \item recursively build measure descriptors for each of the SCCs.
  \item build the directed acyclic graph of SCCs and enumerate the SCCs in the graph.
  \item \texttt{cons} the enumeration of each SCC onto the measure descriptors from the recursive calls.
  \end{enumerate}
\end{itemize}

\begin{figure}[!h]
\scriptsize
\begin{verbatim}
;;                (<abstract node -- reachable bake-rank-map>) . (<measure-descriptor>)
;;                ---------------------------------------------------------------------
((((:LOC 0)  (:DONE NIL) (:LOOP=0 T)   (:RUNS=0 NIL) (:INV T)) . (4 RUNS 11 0))
 (((:LOC 1)  (:DONE NIL) (:LOOP=0 T)   (:RUNS=0 NIL) (:INV T)) . (4 RUNS 10 0))
 (((:LOC 2)  (:DONE NIL) (:LOOP=0 T)   (:RUNS=0 NIL) (:INV T)) . (4 RUNS 9  0))
 (((:LOC 3)  (:DONE NIL) (:LOOP=0 NIL) (:RUNS=0 NIL) (:INV T)) . (4 RUNS 8  LOOP 2 0))
 (((:LOC 4)  (:DONE NIL) (:LOOP=0 NIL) (:RUNS=0 NIL) (:INV T)) . (4 RUNS 8  LOOP 1 0))
 (((:LOC 5)  (:DONE NIL) (:LOOP=0 NIL) (:RUNS=0 NIL) (:INV T)) . (4 RUNS 8  LOOP 3 0))
 (((:LOC 5)  (:DONE NIL) (:LOOP=0 T)   (:RUNS=0 NIL) (:INV T)) . (4 RUNS 7  0))
 (((:LOC 6)  (:DONE NIL) (:LOOP=0 T)   (:RUNS=0 NIL) (:INV T)) . (4 RUNS 6  0))
 (((:LOC 7)  (:DONE NIL) (:LOOP=0 T)   (:RUNS=0 NIL) (:INV T)) . (4 RUNS 5  0))
 (((:LOC 8)  (:DONE NIL) (:LOOP=0 NIL) (:RUNS=0 NIL) (:INV T)) . (4 RUNS 4  LOOP 4 0))
 (((:LOC 9)  (:DONE NIL) (:LOOP=0 NIL) (:RUNS=0 NIL) (:INV T)) . (4 RUNS 4  LOOP 3 0))
 (((:LOC 10) (:DONE NIL) (:LOOP=0 NIL) (:RUNS=0 NIL) (:INV T)) . (4 RUNS 4  LOOP 2 0))
 (((:LOC 11) (:DONE NIL) (:LOOP=0 NIL) (:RUNS=0 NIL) (:INV T)) . (4 RUNS 4  LOOP 1 0))
 (((:LOC 12) (:DONE NIL) (:LOOP=0 NIL) (:RUNS=0 NIL) (:INV T)) . (4 RUNS 4  LOOP 5 0))
 (((:LOC 12) (:DONE NIL) (:LOOP=0 T)   (:RUNS=0 NIL) (:INV T)) . (4 RUNS 3  0))
 (((:LOC 13) (:DONE NIL) (:LOOP=0 T)   (:RUNS=0 NIL) (:INV T)) . (4 RUNS 2  0))
 (((:LOC 14) (:DONE NIL) (:LOOP=0 T)   (:RUNS=0 NIL) (:INV T)) . (4 RUNS 1  0))
 (((:LOC 15) (:DONE NIL) (:LOOP=0 T)   (:RUNS=0 NIL) (:INV T)) . (4 RUNS 12 0))
 (((:LOC 15) (:DONE NIL) (:LOOP=0 T)   (:RUNS=0 T)   (:INV T)) . (3 0))
 (((:LOC 16) (:DONE NIL) (:LOOP=0 T)   (:RUNS=0 T)   (:INV T)) . (2 0))
 (((:LOC 17) (:DONE T)   (:LOOP=0 T)   (:RUNS=0 T)   (:INV T)) . (1 0)))
\end{verbatim}
\normalsize
\caption{Result of computing measure descriptors for \texttt{bake-rank-map}}
\label{fig:msrdescr}
\end{figure}

Returning to the Bakery algorithm example, the resulting alist associating abstract graph nodes to
measure descriptors produced by this algorithm is given in Figure~\ref{fig:msrdescr}. The measure
descriptors follow the control flow through the locations of the bakery process state. The outer
loop forms an SCC with the \texttt{runs} component measure breaking the arc from location $14$ to
$15$. Within the outer loop, the inner loops form SCCs with the \texttt{loop} component measure
decreasing to break the SCCs further. The measure descriptor that results allows us to build a
measure showing that the relation \texttt{(and (equal y (bake-tr-next x sh)) (not (bake-tr-done x)))}
is well-founded. In the next section, we cover the theory and its instantiation which allows the
transfer of these measure descriptor results into actual proofs of well-founded relations.

\section{Proving the Generated Measures and Wrapping Up} \label{sec:proof}

\begin{figure}[!htbp]
\scriptsize
\begin{verbatim}
(encapsulate
  (((test-rel-p * *)         => *)
   ((test-map-e *)           => *)
   ((test-map-o * *)         => *)
   ((test-o-bnd)             => *)
   ((test-a-dom)             => *)
   ((test-nexts *)           => *)
   ((test-chk-ord-arc * * *) => *))

  ... <local defuns> ...

  (def-valid-wf-corr-assumption test) ;; macro generating assumptions with "test-" prefix
)
(def-valid-wf-corr-conclusion test)   ;; macro generating derivations with "test-" prefix
\end{verbatim}
\normalsize
\caption{``Theory'' for proving generated measures}
\label{fig:wfothry}
\end{figure}

In the book \texttt{wfo-thry.lisp} from the supporting materials, a ``theory'' is developed
connecting the successful computation of a measure descriptor from the abstract graph to the
definition of a measure proving that the concrete relation is well-founded. The structure of this
book is essentially the definition of two macros --- one macro codifies the assumptions made of a
set of definitions and a second macro generates the conclusions and results derived from these
assumptions. Each macro takes a name prefix parameter which is prepended to all of the definition
and theorem names generated in the macro. \begin{footnote}{It is worth noting that this would be
    better carried out with functional instantiation in ACL2, but due to technical issues, that has
    not worked in all cases --- we are working on rectifying this.}\end{footnote} The end of the
\texttt{wfo-thry} book concludes with the forms in Figure~\ref{fig:wfothry}. The function
\texttt{(rel-p x y)} is the relation we want to prove is well-founded. The function \texttt{(map-e
  x)} is essentially equivalent to the \texttt{bake-rank-map} function from our example and
\texttt{(map-o x o)} is the \texttt{bake-rank-ord} function. The constant \texttt{(a-dom)} is the
set of nodes in the abstract graph (as in Figure~\ref{fig:reachnodes}) and \texttt{(nexts x)} is a
function taking a node in \texttt{(a-dom)} and returning a list of successor nodes pulled from the
abstract graph. The function \texttt{(chk-ord-arc x y o)} takes two nodes in the abstract graph
\texttt{x} and \texttt{y} and a component measure name \texttt{o} and returns either \texttt{:<<} if
the measure is strictly-decreasing, \texttt{t} if it is non-increasing, and \texttt{nil} if it is
possibly-increasing.

\begin{figure}[!htbp]
\scriptsize
\begin{verbatim}
(defthm map-e-member-nexts
  (implies (rel-p x y)
           (in-p (map-e y) (nexts (map-e x)))))

(defthm map-o-decrement-strict
  (implies (and (rel-p x y)
                (equal (chk-ord-arc (map-e x) (map-e y) o) :<<))
           (bnl< (map-o y o) (map-o x o) (o-bnd))))

(defthm map-o-decrement-non-strict
  (implies (and (rel-p x y)
                (equal (chk-ord-arc (map-e x) (map-e y) o) t))
           (bnl<= (map-o y o) (map-o x o) (o-bnd))))
\end{verbatim}
\normalsize
\caption{Assumptions for proving generated measures}
\label{fig:msrassume}
\end{figure}

The main theorems assumed about these functions are provided in Figure~\ref{fig:msrassume}. These
assumed properties provide the correlation between \texttt{(rel-p x y)} and the abstract graph
defined by \texttt{(nexts x)} with the tagging of component measure ordering on the arcs defined by
\texttt{chk-ord-arc}. The relation \texttt{(bnl< m n o)} is defined in the book \texttt{bounded-ords.lisp}
and orders lists of naturals \texttt{m} and \texttt{n} of length \texttt{o} as the lexicographic
product of the naturals in the list in order. The \texttt{bounded-ords} book defines functions and
relations for building and ordering lists of naturals (of the same length) and lists of natural
lists (potentially differing lengths). These are recognized as \texttt{bnlp} and \texttt{bnllp} and
ordered with \texttt{bnl<} and \texttt{bnll<} respectively. As an aside, we note that in the supporting
books for the paper, we use \texttt{bplp} instead of \texttt{bnlp} where \texttt{bplp} is the same
as \texttt{bnlp} but requires the last natural in the list to not be $0$. The reason to use
\texttt{bplp} is to allow a list of all $0$s to be used as a bottom element in certain
constructions. We use \texttt{bnl} and \texttt{bnll} throughout the paper for clarity.

There are also conversion functions \texttt{bnl->o} and \texttt{bnll->o} for converting
\texttt{bnl}s and \texttt{bnll}s to ACL2 ordinals preserving the well-founded ordering \texttt{o<}
on ACL2 ordinals. We use the bounded ordinals from \texttt{bounded-ords} instead of ACL2 ordinals
because these bounded ordinals are closed under lexicographic product while ACL2 ordinals are
not. This allows us to build constructions using these bounded ordinals that would be far more
difficult (and in some cases, not even possible) with ACL2 ordinals.

\begin{figure}[!htbp]
\scriptsize
\begin{verbatim}
(defthm msr-is-o-p (o-p (msr x m)))
         
(defthm relp-well-founded
  (implies (and (valid-omap m)
                (rel-p x y))
           (o< (msr y m) (msr x m)))

(defthm mk-bnl-is-bnlp
  (implies (valid-omap m)
           (bnlp (mk-bnl x m) (bnl-bnd m)))
         
(defthm mk-bnl-transfers-rel-p-bnl<
  (implies (and (valid-omap m)
                (rel-p x y))
           (bnl< (mk-bnl y m) (mk-bnl x m) (bnl-bnd m)))
\end{verbatim}
\normalsize
\caption{Derivations for proving generated measures}
\label{fig:msrderive}
\end{figure}

Given the assumptions from Figure~\ref{fig:msrassume} (and some additional typing assumptions), we
generate a measure function \texttt{(msr x m)} which takes an ACL2 object \texttt{x} and a measure
descriptor mapping produced from the SCC decomposition (as in Figure~\ref{fig:msrdescr} and termed
an \texttt{omap}) and produces an ACL2 ordinal. If the mapping \texttt{m} satisfies the generated check
\texttt{(valid-omap m)}, then the \texttt{(msr x m)} returns a strictly decreasing ordinal which shows
that \texttt{(rel-p x y)} is well-founded. The key derived properties generated from the instantiation
of this theory are provided in Figure~\ref{fig:msrderive}. In addition to generating the definition
of a measure function returning ACL2 ordinals, we also generate a definition producing a bounded
ordinal \texttt{bnlp} and related properties. The intent is to use the \texttt{mk-bnl} function when one
wants to use the generated ordinal in a composition to build larger ordinals --- even potentially
using the procedure in this paper hierarchically where the component measure at one level is a proven
generated measure at a lower level in the hierarchy.

Returning to the Bakery example, the book \texttt{bake-proofs} includes the generated abstract
graphs and measure descriptor mappings (or \texttt{omap}s) from \texttt{bake-models} and sets up
instantiations of the ``theory'' from the \texttt{wfo-thry} book. The result is the generated
measures for proving our target two relations are well-founded: the relation defined by the step
function \texttt{bake-tr-next} and the blocking relation \texttt{bake-tr-blok}. In the book
\texttt{top.lisp}, we use these results to reach our goal of defining and admitting the
\texttt{bake-run} function from Figure~\ref{fig:bakerun}. We use the generated measure function
\texttt{bank-rank-mk-bnl} to define the function \texttt{(bake-rank-bnll l sh)} which conses the
\texttt{bnl}s for each bakery process state in the list \texttt{l} and returns a \texttt{bnll}. The
measure we use for admitting \texttt{(bake-run st orcl)} is the conversion of the resulting {\tt
  bnll} to ACL2 ordinals:

\begin{verbatim}
           (bnll->o (len (bake-st->trs st))
                    (bake-rank-bnll (bake-st->trs st)
                                    (bake-st->sh st))
                    (bake-rank-bnl-bnd))
\end{verbatim}
\normalsize

Additionally, as we noted earlier in Section~\ref{sec:bakery}, we need the generated measure for
proving\\
\texttt{bake-tr-blok} is well-founded in order to define \texttt{choose-ready} correctly. In
particular,\\
\texttt{(choose-ready l sh o)} is a constrained function which ensures that if there is
a bakery state which is not done in \texttt{l}, then \texttt{(choose-ready l sh o)} will return the
index of a state which is not done and not blocked. The local witness in the encapsulation for
\texttt{choose-ready} is:

\begin{verbatim}
           (local (defun choose-ready (l sh o)
                    (find-unblok (find-undone l) l sh)))
\end{verbatim}
\normalsize

Where \texttt{(find-undone l)} returns an index in \texttt{l} for a bakery process which is not done
(if one exists) and the function \texttt{(find-unblok n l sh)} takes an index \texttt{n} and finds
an index which is not blocked in \texttt{l}. The function \texttt{find-unblok} is (essentially)
defined as:

\begin{verbatim}
           (define find-unblok ((n natp) (l bake-tr-lst-p) (sh bake-sh-p))
             (if (bake-blok (nth n l) l)
                 (find-unblok (pick-blok (nth n l) l) l sh)
               n))
\end{verbatim}
\normalsize

Where \texttt{(bake-blok a l)} returns true if any state in \texttt{l} blocks \texttt{a} and
\texttt{(pick-blok a l)} finds an index in \texttt{l} for a bakery state which blocks \texttt{a} (and
thus if \texttt{(bake-blok a l)} then \texttt{(bake-tr-blok a (nth (pick-blok a l) l))}). The measure
used to admit \texttt{find-unblok} is defined using the generated measure \texttt{bake-nlock-msr} for
proving the \texttt{bake-tr-blok} relation is well-founded.

An additional important property of \texttt{(find-unblok n l sh)} is that it either returns \texttt{n} (if
\texttt{(nth n l)} is not blocked) or it returns an index for a process state which is blocking another
state. We prove separately that no process in a done state can block another process and thus, if
the state at index \texttt{n} passed to \texttt{find-unblok} is not in a done state, then the state at the
index returned by \texttt{find-unblok} is also not in a done state.

\section{Related Work and Future Work} \label{sec:concl}

The analysis of abstract reachable graphs with ordering tags is similar in some ways to analysis of
automata on infinite words~\cite{buchi}, but our search for a measure construction is not the same as
language emptiness or other checks typically considered for infinite word or tree automata. As we
mentioned in Section~\ref{sec:intro}, our work does share similarity to the work on CCGs in ACL2s in
that both build and analyze graphs with the goal of showing that no ``bad'' infinite paths exist
through the graph --- but the focus and approach of each work is significantly different. CCG
termination analysis in ACL2s is used to determine if a user-specified function always terminates. A
significant component of the CCG analysis is unwinding and transforming CCGs (and CCMs) until one
can no longer find bad paths and thus ensure termination. The problem that CCG analysis attempts to
tackle is intrinsically more difficult than the problems we target. We attempt to prove a given
relation is well-founded and rely on mapping function definitions to build a closed model sufficient
to then find a proven measure. Our procedure aggressively builds the model as specified by the
mapping functions and proceeds assuming it is sufficient without further refinement -- the user or
outside heuristics are responsible for any further refinements. The AProVE program analysis
tool~\cite{aprove} provides mechanisms for automatically checking program termination. The approach
taken in AProVE is to translate the source program into a term rewriting system and apply a
variety of analysis engines from direct analysis of the term rewriting to translation into checks
for SAT or SMT. Similar to CCG analysis in ACL2s, the primary difference between AProVE and our work
is an issue of focus. The contexts we target benefit from the assumption of mapping functions and
(in the case of this paper) finite-state systems which can be processed by {\tt GL}. This allows us
to build the tagged abstract reachable graph directly with less reliance on having sufficient
rewrite rules and term-level analysis and heuristics.

There are many ways to extend the work presented. We would like to add an interface into
\texttt{SMTLINK} \cite{smtlink} for either building the abstract graphs and/or the addition of
ordering tags to the arcs in the graphs. \texttt{SMTLINK} is more limited than \texttt{GL} in what
ACL2 definitions it can support, but \texttt{SMTLINK} would be a nice option to have in the cases
where the definitions were viable for \texttt{SMTLINK}. While we assumed the definition of mapping
functions and component measures for the sake of this paper, it is not difficult to write heuristics
to generate candidate mappings and measures either from datatype specifications in the source
definition or from static analysis results. Further, the results and steps in the process can be
analyzed to determine which refinements to apply to the mapping and component measures in an ``outer
loop'' to our procedure. From our limited experience, this is best addressed with guidance from the
domain of application and the types of relations and systems that the user wants to analyze.

\newpage



\nocite{*}
\bibliographystyle{eptcs}
\bibliography{mybib}
\end{document}